\documentclass[aps,prl,amsmath,amssymb,superscriptaddress,notitlepage,groupedaddress]{revtex4-1}
\pdfoutput=1
\usepackage{graphicx}
\usepackage{dcolumn}
\usepackage{stackengine}
\usepackage[section]{placeins}
\usepackage[dvipsnames*,svgnames]{xcolor}
\usepackage{verbatim}
\usepackage[position=bottom,caption=false,captionskip=3pt,font=normalsize,subrefformat=simple,labelformat=simple]{subfig}

\usepackage{listings}
\usepackage{mathtools}
\usepackage{amsmath}
\usepackage{chngcntr}
\usepackage{bm}
\usepackage{url}

\usepackage[breaklinks]{hyperref}
\usepackage{tabularx}
\usepackage{soul}
\hypersetup{
colorlinks,
linkcolor={blue!100!black},
citecolor={blue},
urlcolor={blue!80!black}}
\def\equationautorefname~#1\null{#1\null}
\usepackage[bottom]{footmisc} \makeatletter \def\p@section{} \def\p@subsubsection{} \makeatother
\begin{document}

\title{Bayesim: a tool for adaptive grid model fitting with Bayesian inference}

\author{Rachel Kurchin}
\email{rachel.kurchin@gmail.com}
\affiliation{Massachusetts Institute of Technology, 77 Massachusetts Avenue, Cambridge, MA 02139, USA}
\author{Giuseppe Romano}
\affiliation{Massachusetts Institute of Technology, 77 Massachusetts Avenue, Cambridge, MA 02139, USA}
\author{Tonio Buonassisi}
\email{buonassi@mit.edu}
\affiliation{Massachusetts Institute of Technology, 77 Massachusetts Avenue, Cambridge, MA 02139, USA}

\begin{abstract}
  Bayesian inference is a widely used and powerful analytical technique in fields such as astronomy and particle physics but has historically been underutilized in some other disciplines including semiconductor devices. In this work, we introduce \texttt{bayesim}, a Python package that utilizes adaptive grid sampling to efficiently generate a probability distribution over multiple input parameters to a forward model using a collection of experimental measurements. We discuss the implementation choices made in the code, showcase two examples in photovoltaics, and discuss general prerequisites for the approach to apply to other systems.
\end{abstract}

\maketitle

\section*{Introduction}
  There are a plethora of examples across diverse scientific and engineering fields of mathematical models used to simulate the results of experimental observations. In many cases, there are input parameters to these models which are difficult to determine via direct measurement, and it is desirable to invert the numerical model -- that is, use the experimental observations to determine values of the input parameters. Bayesian inference, or Bayesian parameter estimation (BPE) is a fruitful framework within which to do such fitting, since the resulting posterior probability distribution over the parameters of interest can give rich insights into not just the most likely values of the parameters, but also uncertainty about these values and the potentially complicated ways in which they can covary to give equally good fits to observations. It has been widely used for decades in fields such as cosmology~\cite{cosmobayesreview}, particle physics~\cite{particlebayes, higgsbayes}, and biology~\cite{evobiobayes, sysbiobayes}, with well-developed and widely used code packages specialized to these purposes.~\cite{multinest, biobayescode} However, the use of Bayesian inference in materials science and related contexts is much more recent, having only emerged in the past few years.~\cite{lookman, combo, SnSJoule}

  We have previously demonstrated the value of a BPE approach in using automated high-throughput temperature- and illumination-dependent current-voltage measurements ($JVTi$) to fit material/interface properties and defect recombination parameters in photovoltaic (PV) absorbers~\cite{SnSJoule,FeBayes}, i.e. as a replacement for direct experimental characterization via probabilitistic inversion of a forward model. In cases such as these, when the data model is not a simple analytical equation but rather a computationally intensive numerical model, efficient, nonredundant sampling of the parameter space when computing likelihoods becomes critical to making the fit feasible.

  In this work, we introduce \texttt{bayesim}, a Python-based code that utilizes adaptive grid sampling to perform BPE. It is designed to be easy to use by anyone with basic familiarity with Python. We discuss the structure of the code, its implementation, and provide several examples of its usage. While the authors' expertise is in the realm of semiconductor physics and thus the examples herein are drawn from that space, we also discuss the general characteristics of a problem amenable to this approach so that researchers from other fields might adopt it as well.

\section*{Technical Background}

 Bayes' Theorem is a relationship between conditional probabilities. It states

 \begin{equation}
   \begin{split}
     \label{Eq:1001}
     P(H|E)=\frac{P(H)P(E|H)}{P(E)},
   \end{split}
 \end{equation}

 where the notation $P(A|B)$ indicates the probability of $A$ being true given that $B$ is true. $H$ is a \textit{hypothesis} and $E$ the observed \textit{evidence}. $P(H)$ is termed the \textit{prior}, $P(E|H)$ the \textit{likelihood}, $P(H|E)$ the \textit{posterior}, and $P(E)$ is a normalizing constant. If there are $n$ pieces of evidence, this can generalize to an iterative process where

 \begin{equation}
   \begin{split}
     \label{Eq:2}
     P(H|\{E_1,E_2,...E_n\}) = \frac{P(H|\{E_1,E_2...E_{n-1}\})P(E_n|H)}{P(E_n)},
   \end{split}
 \end{equation}

 i.e. as each new piece of evidence is fed in, the posterior probability distribution is updated to accommodate it and should in general become more concentrated into one region of parameter space. In a multidimensional parameter estimation problem, each hypothesis $H$ is a tuple of possible values for the fitting parameters, i.e. a point in the parameter space, while the evidence $E$ is an observed output of a measurement as a function of various experimental conditions.

 In order to compute a likelihood, a model capable of simulating that observed output as a function of both the fitting parameters and the experimental conditions is required. In \texttt{bayesim}, likelihoods are calculated for each point in parameter space using a Gaussian where the argument is the difference between observed and simulated output at that point, and the standard deviation is the sum of experimental uncertainty and model uncertainty. The experimental uncertainty is a number provided by the user that quantifies any noise/irreproducibility inherent to the measurement, while the model uncertainty is calculated by \texttt{bayesim} and reflects the sparseness of the parameter space grid, i.e. how much simulated output changes from one grid point to another.

 \begin{figure}
   \includegraphics[width=0.9\columnwidth]{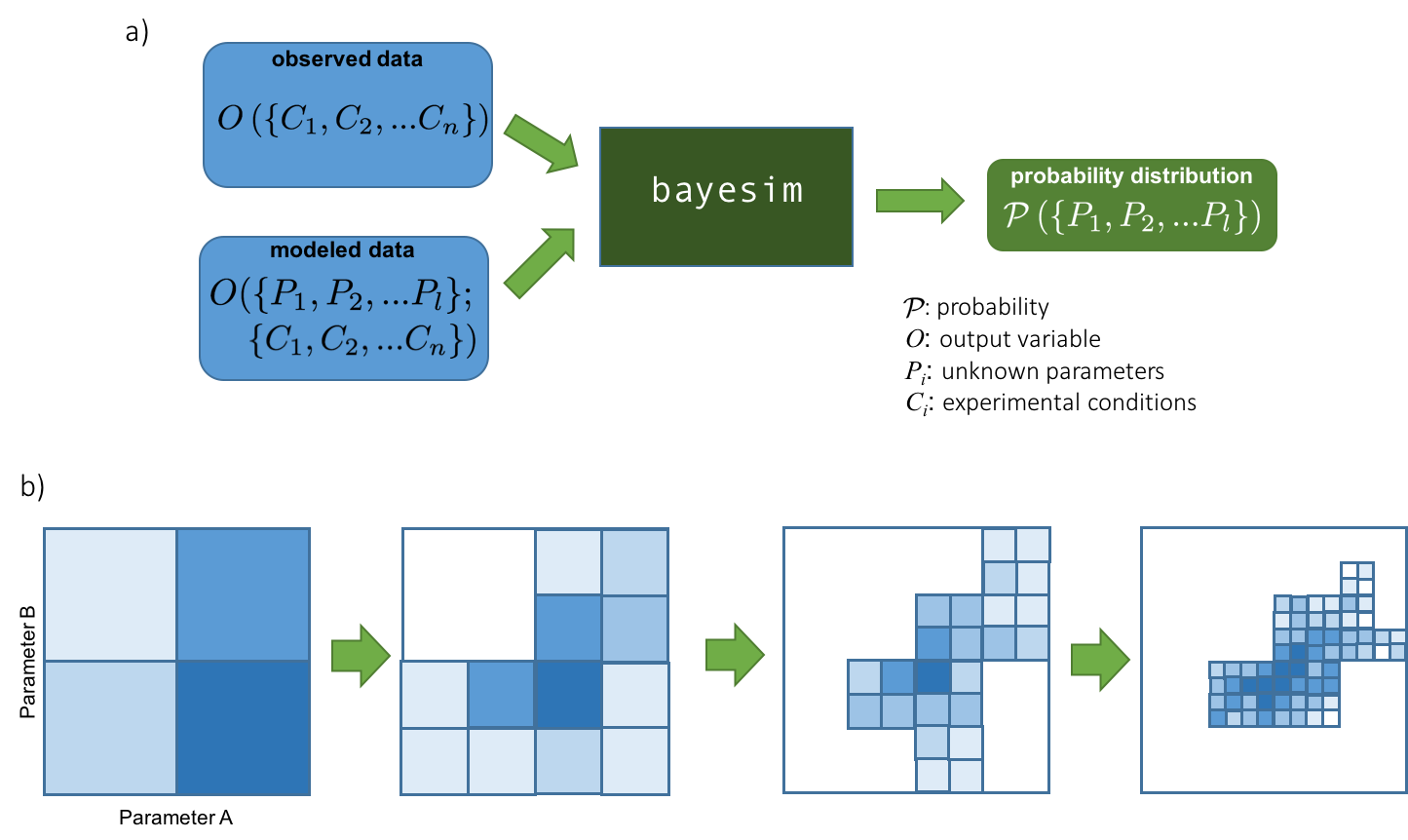}
   \caption{a) High-level flowchart of inputs and outputs to \texttt{bayesim}. Observed data as a function of experimental conditions and modeled data over those same conditions as well as fitting parameters are fed in, and the code produces a probability distribution over the fitting parameter values. b) 2D schematic of adaptive grid sampling approach, illustrating progressive subdivision in higher-probability regions of parameter space.}
   \label{fig1}
 \end{figure}

 A high-level summary of what \texttt{bayesim} does is shown in ~\autoref{fig1}a. The flowchart shows that observed (as a function of experimental conditions $\{C\}$ and simulated (as a function of fitting parameters $\{P\}$ and experimental conditions $\{C\}$) outputs are compared to produce a probability distribution over $\{P\}$. ~\autoref{fig1}b schematically indicates the adaptive grid sampling approach for a hypothetical two-dimensional parameter space, wherein grid boxes exceeding some threshold probability are subdivided and lower-probability regions discarded, allowing attainment of a high fitting precision without needing to sample the entire parameter space at the same high density.

\section*{Software Architecture and Interface}

  \begin{figure}
    \includegraphics[width=0.95\columnwidth]{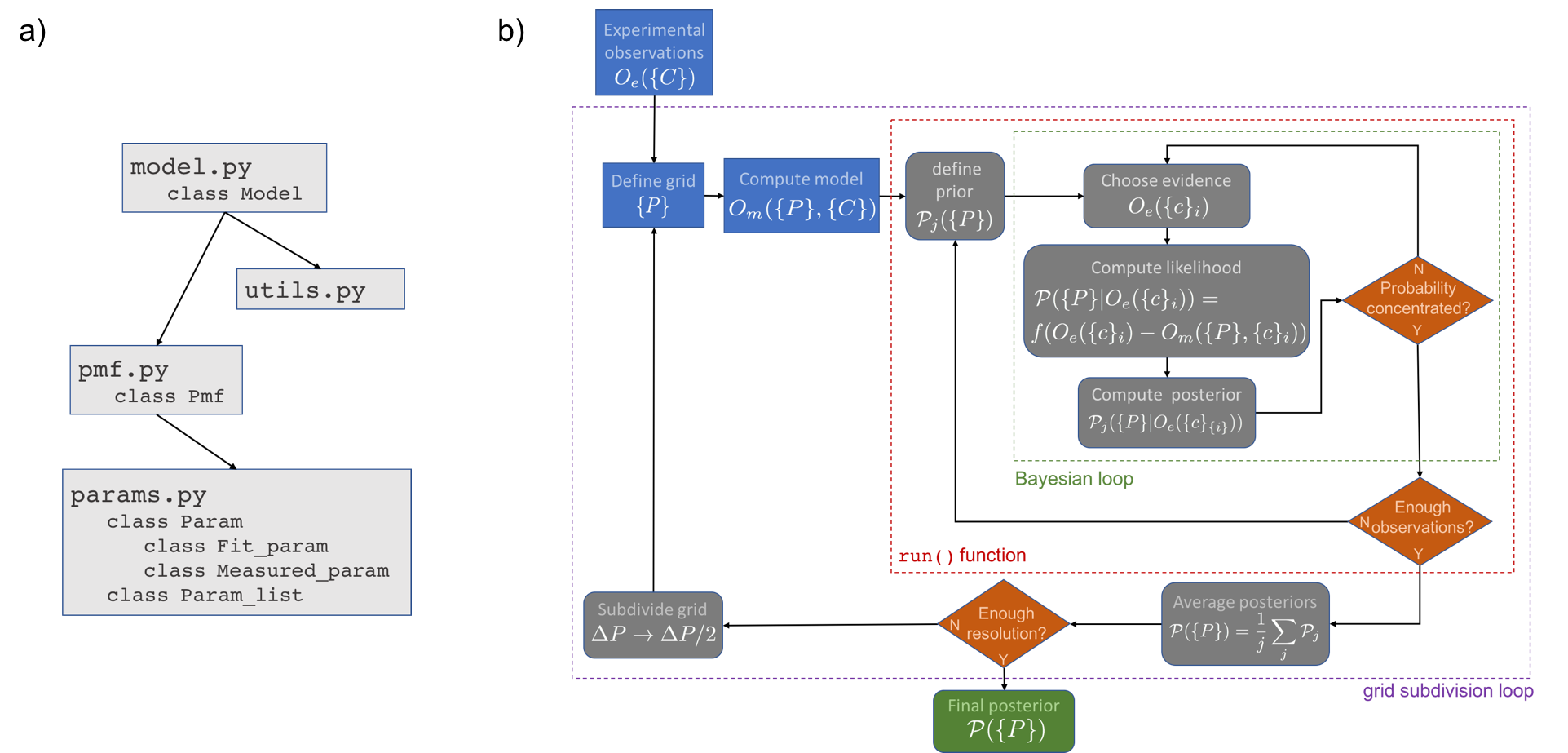}
    \caption{a) \texttt{bayesim} software structure. b) Detailed flowchart of use of \texttt{bayesim}, with calculation steps and various levels of iteration indicated. Boxes in blue indicate steps that can or must be done explicitly by the user, grey steps that are done by the code with behavior tunable by the user, orange diamonds control points for staying in or leaving a loop, and green the final output.}
    \label{structure}
  \end{figure}

  The basic structure of \texttt{bayesim} is shown in ~\autoref{structure}a. Detailed and up-to-date documentation of classes, functions, and example applications is maintained online.~\cite{docs} The top-level object with which users interact is implemented in the \texttt{Model} class. The \texttt{params} module defines classes to store information about the various types of parameters (fitting parameters, experimental conditions, and measured output) while the \texttt{Pmf} class stores the probability distribution and implements the manipulations required for Bayesian updates.

  ~\autoref{structure}b is a flowchart of using the code, with the main inference loop surrounded by a dotted line. One first needs to provide the observed and modeled data. Observed data is accepted in HDF5 format, while modeled data can be an HDF5 file (in which case \texttt{bayesim} will determine the fitting parameter space grid from this file), or if the fitting parameter space has been predefined, one can also attach a Python function that computes the model data (e.g. an interface to a numerical model). Observed data should have associated experimental uncertainties, or the user can indicate a fixed value of uncertainty to use for all observations upon attaching the data.

  Once modeled data is attached, model uncertainty can be computed. At each point in the space of experimental conditions, a matrix of the simulated output variable as a function of the fitting parameters is constructed and at each point in this matrix, the largest difference along any index is computed. This is set as the model uncertainty for this set of experimental conditions at this point in parameter space.

  At this point, the Bayesian inference loop can proceed. We start with a bounded uniform prior distribution over the entire parameter space grid and the observed data in a randomized order. We feed in one piece of evidence (i.e. one observed data point) and compute the likelihood using a Gaussian. The standard deviation for each point in the parameter space is the sum of the experimental uncertainy of that observation and the model uncertainty at that point. The posterior distribution is computed as a Bayesian update on the prior. Next, we check whether the threshold condition for concentration of probability (defined as at least some percentage (by default 90\%) of the probability mass residing in at most some percentage (by default 5\%) of the parameter space volume. If this threshold is satisfied, the current posterior is stored and the prior reset to a uniform distribution. If not, the next observation is fed in. This procedure is repeated until the requisite number (by default, 80\%) of observed data points has been used, with the final posterior being the average of every stored posterior. The posterior is calculated in this way (rather than a Bayesian multiplication through every single observed point) to avoid numerical errors that can arise if very small numbers are multiplied together. This procedure of running the Bayesian loop multiple times until the desired number of pieces of evidence are used is encapsulated in the \texttt{Model.run()} function.

  Now the grid can be subdivided. The user defines a threshold probability (by default 0.001) for a grid box to possess, and all boxes with greater than this probability, as well as all their immediate neighbors, are divided in two along every parameter. \texttt{bayesim} saves a list of the new model calculations that need to be done for inference to proceed again, and the user provides these, again either as a callable or an HDF5 file. For the next round of inference, the user can also define a ``relaxation" parameter from 0 to 1.0 (default 0) that will mix in some fraction of the posterior from the previous step with the uniform distribution over this new grid.

  This overall procedure can be repeated as many times as the user wishes to get a more precise fit -- eventually, the concentration of probability becomes limited by experimental rather than model uncertainty. There is also an option to define a minimum width for a box along any parameter direction, and if a box is already less than this width, \texttt{bayesim} will not subdivide in that direction.

  The most flexible way to interact with \texttt{bayesim} is via Python scripting or through literate programming in a Jupyter notebook. There is also a command line interface (CLI) for users less familiar with coding in Python planned for future release.

  \texttt{bayesim} relies on a variety of external open-source packages. These include \texttt{numpy}~\cite{numpy} and \texttt{scipy}~\cite{scipy} for a variety of mathematical functions and vectorized implementations, \texttt{joblib}~\cite{joblib} for simple parallelism, \texttt{deepdish}~\cite{deepdish} for saving and loading HDF5 files, \texttt{pandas}~\cite{pandas} for data manipulation, and \texttt{matplotlib}~\cite{mpl} for visualization.

\section*{Application Examples}
  \subsection{Ideal Diode Model}
    As a first example, we fit a simple two-parameter model of solar cell current density $J$ (as a function of voltage $V$ and temperature $T$) known as the ideal diode model:

    \begin{equation}
      J(V,T) = J_L+J_0\left(\exp{\frac{qV}{nkT}}-1\right),
      \label{ID_eqn}
    \end{equation}

    where $k=8.61733\times 10^{-5}$ eV/K is Boltzmann's constant, by convention $J_L$ (the light current) is negative and $J_0$ (the saturation current) is positive but strongly dependent on temperature, a dependence we can approximate as:

    \begin{equation}
      J_0 \approx B'T^{3/n}\exp{\left(\frac{-E_{g0}}{nkT}\right)},
    \end{equation}

    where $E_{g0}$ is the zero-temperature bandgap of the material. For this example, we will set this to 1.2 eV and $J_L$ to -0.03 mA/cm$^2$. There are thus two parameters to be fit, the coefficient $B'$ (which can range over a wide variety of values) and the ideality factor $n$ (which is by definition between 1 and 2).

    \begin{figure}
      \includegraphics[width=0.8\columnwidth]{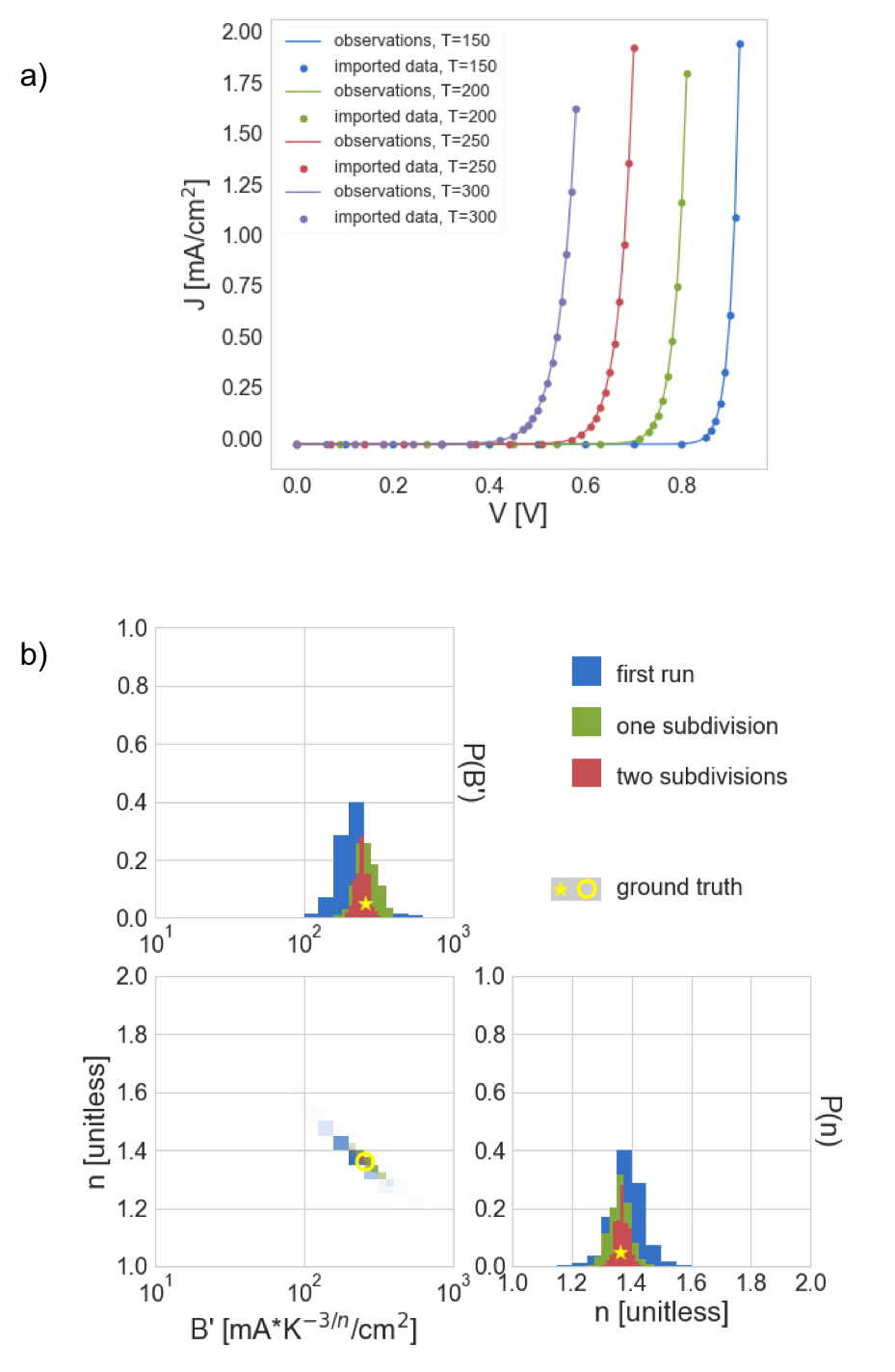}
      \caption{a) ``Observations" generated by ideal diode model, Equation~\ref{ID_eqn}, with overlaid scatter to show which points were actually imported and used by \texttt{bayesim}, b) PMF's after the first inference run, second (one subdivision), and third (two subdivision), in blue, orange, and green, respectively. Values used to generate data indicated in red.}
      \label{ID_data}
    \end{figure}

    As validation, we generate the ``observations" using the same model we'll use for fitting, with parameter values of $n=1.36$ and $B'=258$, and four temperature values of 150, 200, 250, and 300 K. These ``observations" are shown in ~\autoref{ID_data}a. When \texttt{bayesim} imports observed data, it can optionally discard data points that are very close by to each other in order to reduce computational load on modeling less informative points -- the observed points that were actually imported (59 out of 305 total ``observations") are shown as overlaid scatter dots.

    ~\autoref{ID_data}b shows the posterior distributions after the first inference run and after each round of subdivision, demonstrating that \texttt{bayesim} gets progressively closer to the ``true" values as the loop continues iterating. In addition, on the first inference run (prior to any subdivision), an average of 45\% of grid points had larger model uncertainty than experimental uncertainty, while after the second subdivision, only 1.9\% did, while the average absolute error of the highest-probability model decreased from 0.03 after the first run to 0.007 after the second subdivision.

    In addition, the initial grid had 400 boxes, the singly subdivided grid 164, and the final grid (4x initial resolution) had 144, meaning the overall additional computational effort to quadruple the fit precision was less than producing the initial simulation set. In comparison, using a ``brute-force'' grid approach, reaching 4x precision would take $4^n$ times the computational load in $n$ dimensions, i.e. 16x for this case. The exact factor of improvement is problem-dependent, but in all cases the load is significantly less than the exponential scaling of the prior approach.

  \subsection{SnS Solar Cell}
    This example demonstrates a more realistic application of \texttt{bayesim}, i.e. when the model being used to compute likelihoods is more computationally expensive. In this case, we will repeat the analysis from Ref.~\cite{SnSJoule}, using SCAPS~\cite{SCAPS} (a coupled drift-diffusion/Poisson solver) as the model, which takes 10-15 core-seconds for a single J-V sweep.

    \begin{figure}
      \includegraphics[width=0.8\columnwidth]{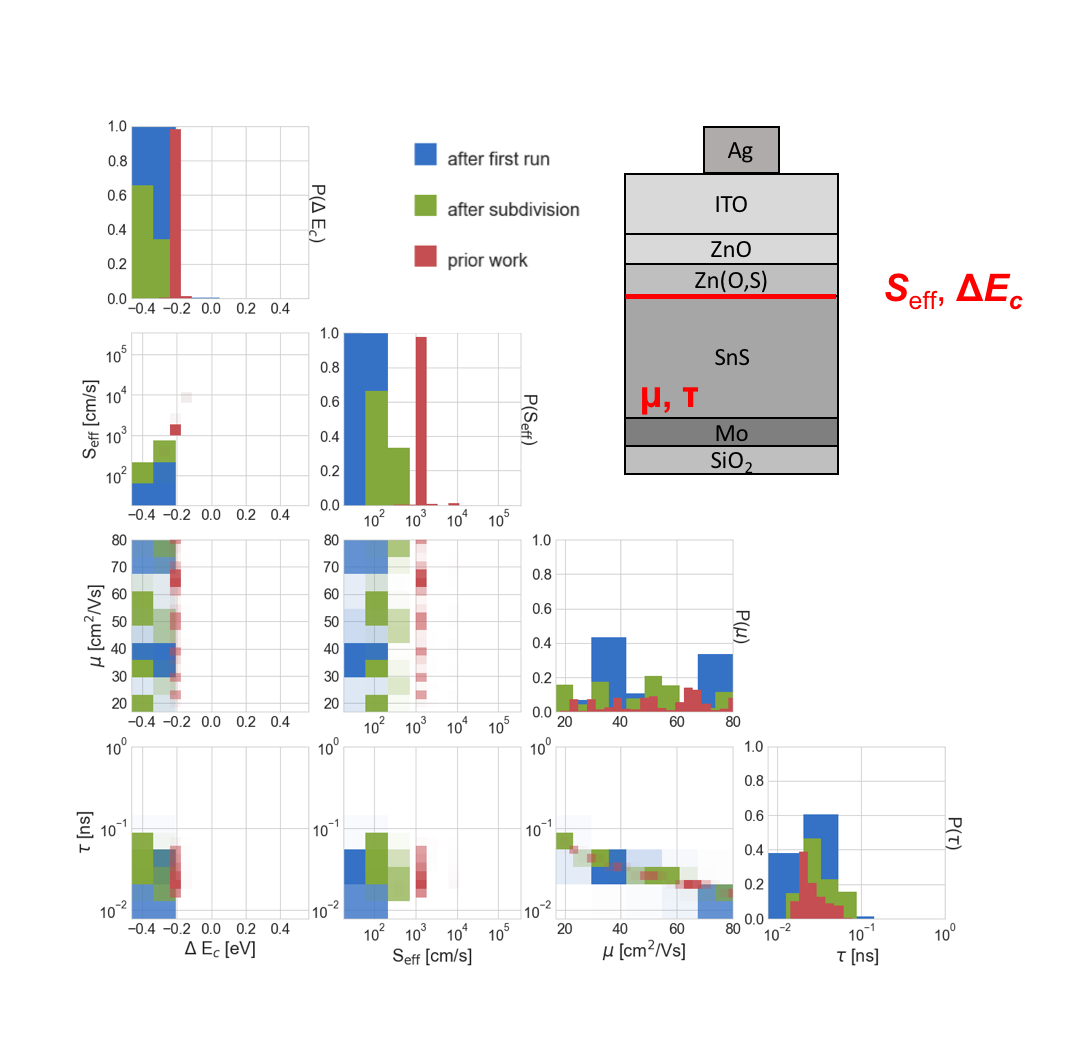}
      \caption{PMFs produced by \texttt{bayesim}, after initial run and after one subdivision (blue and green, respectively). In red, PMF from Reference~\cite{SnSJoule}. Inset: device structure (illuminated from above) with bulk SnS ($\mu$,$\tau$ fit) and SnS/Zn(O,S) interface ($\Delta E_\text{c}$, $S_{\text{eff}}$ fit) indicated.}
      \label{SnS_data}
    \end{figure}

    We start on a significantly coarser grid than Ref.~\cite{SnSJoule} and show only one level of subdivision to make the example feasible to run on a personal computer. \autoref{SnS_data} shows the results from this fit done in \texttt{bayesim} compared to the higher-resolution grid analysis done previously. Note that the inference step in full-resolution grid took a few hours, while the adaptive grid fit takes a few minutes. While the fit precision is of course lower at only one level of subdivision, the extracted values are generally in agreement and more importantly, the relationship between the mobility and lifetime (namely, that the product, i.e. the diffusion length), is constrained, emerges just as clearly. The differences in fitted values stem from the fact that the prior published analysis didn't account for model uncertainty, while this feature is included in \texttt{bayesim}.

\section*{Conclusions}
 We have introduced \texttt{bayesim}, a Python-based adaptive grid sampled Bayesian inference code which is flexible to the model used to calculated likelihood as well as the parameter space over which to fit. We provided two examples of application of the code in the space of characterizing PV devices using JVT(i) measurements, as this is the authors' area of expertise. However, nothing about the code or the approach limits it to application in PV materials. Any situation where there is:
 \begin{itemize}
   \item a forward model that captures the relevant physics (particularly one that is relatively expensive to evaluate), and
   \item a set of measurements of the output of that model as a function of several experimental conditions
 \end{itemize}
 would be amenable to \texttt{bayesim}. The online documentation page~\cite{docs} maintains a list of examples which will grow as the authors become aware of/implement more. The latest version of the code can always be found in the Github repository~\cite{gh}, and the most recent stable and thoroughly-tested version is installable via PyPI~\cite{pypi}. The online documentation also hosts other useful information, such as detailed documentation of every class and function in \texttt{bayesim}, a more extensive technical background section than is included here, a list of planned future features, and an updated manual for usage of the code.

\section*{Acknowledgements}
R.C.K. acknowledges the funding of a Blue Waters Graduate Fellowship. This work was supported by a TOTAL SA research grant funded through MIT\text{ei}, the MIT Energy Initiative.

\bibliography{biblio}
\bibliographystyle{plainurl}

\end{document}